\documentclass{docclass}

\usepackage{epsfig}
\usepackage{amsfonts}
\usepackage{amssymb}
\usepackage{mathrsfs}
\usepackage{theorem}
\usepackage{amsmath}

\usepackage{ifpdf}
\ifpdf
\usepackage{epstopdf}   
\fi

\newcommand{\ket}[1]{\ensuremath{\vert#1\rangle}}
\newcommand{\bra}[1]{\ensuremath{\langle #1\vert}}
\newcommand{\bk}[2]{\ensuremath{\langle #1\vert #2\rangle}}
\newcommand{\kb}[2]{\ensuremath{\vert #1 \rangle \langle #2 \vert}}

\renewcommand{\vec}[1]{\ensuremath{\mathbf{#1}}}

\newcommand{\tr}{\mathrm{tr}}

\def\unity{\mbox{\small 1} \!\! \mbox{1}}

\def\unity{\mbox{\small 1} \!\! \mbox{1}}

\begin{document}
\title{On the Structure of Protocols\\ for Magic State Distillation}
\author{Earl T. Campbell \inst{1,2} \and 
        Dan E. Browne \inst{1,3} }
\institute{Department of Physics and Astronomy, University College London, Gower Street, London, WC1E 6BT, UK. \and \email{e.campbell@ucl.ac.uk} \and \email{d.browne@ucl.ac.uk} }

\maketitle      
\begin{abstract}
We present a theorem that shows that all useful protocols for magic state distillation output states with a fidelity that is upper-bounded by those generated by a much smaller class of protocols.  This reduced class consists of the protocols where multiple copies of a state are projected onto a stabilizer  codespace and the logical qubit is then decoded.  
 \end{abstract}

\section{Introduction}

To prevent noise and decoherence from destroying quantum information,  the information can be stored in delocalised degrees of freedom of a larger system.   This may be an encoding in a stabilizer code, where information is stored in a subspace of states that are eigenstates of particular tensor products of Pauli operators, known as the stabilizer of the code \cite{Steane95,G01a}.  More exotic physical systems, composed of particles obeying anyonic statistics, can store quantum information in topological degrees of freedom \cite{Kit03,RHG01a}.  However, to perform computational tasks it is necessary to   manipulate the quantum information without producing correlated and uncorrectable errors,  which is a more difficult task once the information is delocalized.  

A logical operation that is performed \textit{transversally} does not couple subsystems within the same encoding block \cite{NC01b}. It cannot produce correlated errors, and hence is well suited for a fault tolerant computation scheme.  For example, an encoded Hadamard applied to the 7-qubit Steane code \cite{Steane95} is simply 7 single qubit Hadamards, $H_{L}=H^{\otimes 7}$, and so is transversal. 

It has, however, been recently shown that no stabilizer code can protect against a generic noise model and provide a universal set of transversal gates \cite{Eastin09}.  Furthermore, the only experimentally observed anyons also fall short of providing a universal set of topologically protected gates \cite{Moore90,Lloyd02,Doucot02}.  In stabilizer codes the most common group of transversal gates is the Clifford group, the group of unitaries that map the set of tensor products of Pauli operators to itself.   The Clifford group is also inherently protected in some topological schemes for quantum computation.  Above fault tolerance error thresholds, high fidelity Clifford group unitaries can be achieved. It is therefore well motivated to abstract away from fault tolerant operations on encoded qubits, and consider  ideal Clifford group operations acting on unencoded qubits. 

Quantum devices which only prepare stabilizer states and perform unitaries in the Clifford group can be efficiently simulated by a classical computer, and so cannot offer any computational advantage \cite{G02a,AB01a}.  Despite this, such devices can be promoted to a fully universal quantum computer by using a resource of pure single-qubit non-stabilizer states, such as the so-called magic states~\cite{BraKit05}.  Since preparation of these resource states is not typically fault tolerant, Bravyi and Kitaev introduced ``magic state distillation'', a class of protocols which, given perfect Clifford operations, allow one to distill from many copies of certain mixed states, a state which is closer in fidelity to a pure non-stabilizer state. Repeated iteration of the protocol generates, in the limit of many iterations, one of a number of pure non-stabilizer states, known as the ``magic states''. Given a  supply of such states, Clifford group unitaries are sufficient for universal quantum computation~\cite{BraKit05}. 

A number of protocols for magic state distillation have been proposed \cite{BraKit05,Rei01a,Rei02a,Rei03a}, but all follow the same format.  They prescribe projecting several copies of the initial state onto a stabilizer codespace and then decoding from the codespace to a single qubit state. The prominence of stabilizer codes in magic state distillation, and a dearth of other species of protocol, is rather surprising, since the protocols have several key differences from quantum error correction schemes. For example, the initial state is an uncorrelated tensor product, not a code-state subject to local noise, and unlike in quantum error correction, error syndromes detected via the stabilizer measurements cannot be corrected.  Here we resolve this puzzle by proving that for every effective magic state protocol there exists a stabilizer code protocol that achieves the same or better fidelity. 

In the following sections we first define some key concepts and present examples. In section \ref{main} we present the main theorem of this paper, and in the following sections and appendix we present a proof.
 

\section{Clifford Reductions}


We pose the problem of magic state distillation within the framework of what we call $n$-to-$1$ Clifford reductions:
\begin{definition}  An \textbf{n-to-1 qubit Clifford reduction} takes an $n$-qubit resource state  $\rho_{\mathrm{nR}} $ and outputs a single qubit $\rho'$ using ideal Clifford unitaries, preparation of stabilizer states, classical feedforward, classical randomness, Pauli measurements and postselection.
\end{definition}
Most current protocols for magic state distillation take a resource state composed of $n$ copies of a non-stabilizer state, so $\rho_{\mathrm{nR}}=\rho_{\mathrm{1R}}^{\otimes n} $.  However, our definition encompasses any $n$-qubit mixed state.  This definition covers all possible protocols that can be executed by a device capable of performing Clifford group operations to a quantum system whilst being controlled by a classical computer.  We call a Clifford reduction a successful round of magic state distillation, when the output qubit has an improved fidelity w.r.t some pure non-stabilizer state.  Here we are only really concerned with distillation on the level of an individual round.  However, it is worth noting that generally distillation protocols require many rounds of concatenation, with many copies of the output qubit forming the resource for distillation in the next Clifford reduction.

Formally,  $n$-to-$1$ qubit Clifford reductions are described by quantum operations of the following form:
\begin{equation}
\label{eqn:WholeOp}
 \rho' = \frac{\tr_{\mathrm{nR-1},\mathrm{A}} \left(  \sum_{i} K_{i} \left( \rho_{\mathrm{nR}} \otimes  \kb{0}{0}^{\otimes m} \right) K_{i}^{\dagger} \right)}{\tr \left(  \sum_{i} K_{i} \left( \rho_{\mathrm{nR}} \otimes  \kb{0}{0}^{\otimes m}\right) K_{i}^{\dagger} \right)} \enspace , 
\end{equation}
where the first $n$ qubits are a resource in a non-stabilizer state $\rho_{\mathrm{nR}} $  and the next $m$ qubits are ancilla in the stabilizer state $\ket{0}\bra{0}$.   The partial trace is subscripted by $\mathrm{nR}-1, \mathrm{A}$, to indicate that we trace over the ancillary Hilbert space, collectively labeled $\mathrm{A}$, and every qubit in the resource Hilbert space, labeled $\mathrm{nR}$, except for a single qubit labeled qubit 1.   Each of the $K_{i}$  are Clifford group Kraus operators and have the form:
\begin{equation}
\label{eqn:CliffordKraus}
	K_{i} = k_{i} P_{i,N} C_{i,N} ...  P_{i,x}C_{i,x}... P_{i,1} C_{i,1} \enspace ,
\end{equation}
where $C_{i,x}$ are Clifford group unitaries and $P_{i,x}$ are Pauli projectors.   In terms of the protocol, Pauli projectors occur when we measure a Pauli operator $s_{i,x}$ and postselect on outcome ``+1", or equivalently measure $-s_{i,x}$ and postselect on outcome ``-1", such that $P_{i,x}=(\unity+s_{i,x})/2$.  Finally, $k_{i}$ is a real number smaller than or equal to unity.  As with all quantum operations, we require $\sum_{i}K_{i}^{\dagger} K_{i}\leq \unity$.  The Kraus operator index $i$ labels each \textit{branch} of the protocol,  with proliferation of branches occurring whenever a protocol specifies: (\textit{i}) that we apply a different operation depending on the value of a random classical variable; (\textit{ii}) we make a Pauli measurement but postselect on more than one outcome, possibly feeding forwarding that outcome to determine later operations, or (\textit{iii}) we introduce an ancillary stabilizer state that is mixed.  We can assume the ancilla are in the state $\kb{0}{0}^{\otimes m}$ without loss of generality, as a Kraus operator can always rotate this state into a different pure stabilizer state.  

\section{The Steane Code Protocol}

\begin{figure}[t]
\centering
\includegraphics[width=\columnwidth]{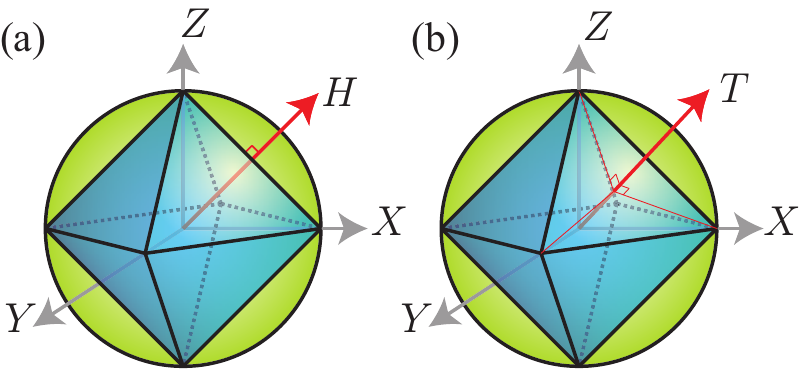}
\caption{The octahedron of mixed stabilizer states within the Bloch sphere with (a) the axis of the Hadamard rotation shown (b) the axis of the T rotation shown.  The Hadamard axis bisects an edge of the stabilizer octahedron, and the T axis bisects the center of a face of the stabilizer octahedron.  About their axes the Hadamard rotation is a 180 degree rotation, and the T rotation is a 120 degree rotation.  Notice that the stabilizer octahedron has six vertices corresponding to the six pure stabilizer states.}
\label{fig:Outline}
\end{figure}

Before continuing with our general examination of protocols for magic state distillation, we will give a concrete example.  The single-qubit pure stabilizer states are eigenstates of $X$, $Y$ or $Z$ and mixed stabilizer states are any probabilistic ensembles of these pure states.   In the Bloch sphere, this convex set with 6 vertices forms the \textit{stabilizer octahedron} shown in figure~\ref{fig:Outline}.  The only non-stabilizer states known to be distillable from mixed states are eigenstates of Clifford group unitaries, such as the Hadamard $H$ and the $T$ operation\footnote{The T rotation performs, $TXT^{\dagger}=Y$, $TYT^{\dagger}=Z$.}.  The corresponding eigenstates are denoted $\ket{H}=H\ket{H}$ and $\ket{T}=T\ket{T}$, and depolarized mixtures for these eigenstates are:
\begin{eqnarray}
\rho(f, \sigma_{H})& = & ( \unity + (2f-1)  (X+Z) / \sqrt{2}   ) / 2 \enspace , \\ \nonumber
\rho(f, \sigma_{T})& = & ( \unity + (2f-1) (X+Y+Z) / \sqrt{3} ) / 2 \enspace ,
\end{eqnarray}
with fidelity $f=1,0$ for ideal magic states and $0<f<1$ when noisy. Other distillable magic states emerge from symmetries the stabilizer octahedron. Given the ability to prepare noisy copies of $\rho(f, \sigma_{H})$, we can attempt a protocol proposed by Reichardt \cite{Rei01a}:
\begin{enumerate}
  \item Prepare a resource state $\rho_{\mathrm{nR}}=\rho(f,\sigma_{T})^{\otimes 7}$;
  \item Measure the 6 stabilizer generators of the Steane code;
  \item If any of the measurements give $-1$ then restart;
  \item Otherwise, the protocol has succeeded. Decode  the encoded state  from the Steane codespace to a single qubit state.
\end{enumerate}
Provided the fidelity,  with respect to $\ket{T}$, of the initial noisy copies exceeds some threshold, the output qubit has an improved fidelity.  Concatenation allows improvement towards fidelities arbitrarily close to unity.  For completeness, we list the stabilizer generators of the Steane code:
\begin{eqnarray}
	X_{1}X_{2}X_{3}X_{4}\unity_{5}\unity_{6}\unity_{7}  ,& X_{1}X_{2}\unity_{3}\unity_{4}X_{5}X_{6}\unity_{7} ,& X_{1}\unity_{2}X_{3}\unity_{4}X_{5}\unity_{6}X_{7} \enspace ,\\ \nonumber
	Z_{1}Z_{2}Z_{3}Z_{4}\unity_{5}\unity_{6}\unity_{7}  ,& Z_{1}Z_{2}\unity_{3}\unity_{4}Z_{5}Z_{6}\unity_{7} ,& Z_{1}\unity_{2}Z_{3}\unity_{4}Z_{5}\unity_{6}Z_{7} \enspace .
\end{eqnarray}
Note that the sign of all these operators is positive. Also note that the protocol post-selects only states resulting from positive measurement outcomes, and that states from other outcomes are discarded. As we will see later, in magic state distillation protocols there is always one choice of measurement outcome that leads to a fidelity unsurpassed by any other outcome.

\section{Stabilizer Reductions}

In addition to Reichardt's protocol their are numerous other protocols for magic state distillation, but they follow the same pattern of projecting onto the codespace of a stabilizer code, and decoding the logical qubit onto a single qubit.  Codes that have been used in this context include the Steane code \cite{Rei01a}, the 5-qubit code \cite{BraKit05}, the 23 qubit Golay code \cite{Rei01a}, a 15-qubit quantum Reed-Muller code \cite{BraKit05}, and the simple 2 qubit parity checking code \cite{Rei03a}.  Furthermore, numerous other protocols derived from stabilizer codes have been numerically tested \cite{BraKit05,Rei03a}.  Note that, Knill's scheme for postselected quantum computing \cite{Knill05} contains a protocol reminiscent of magic state distillation, but which at first appears to be unrelated to any stabilizer code.  However, Reichardt \cite{Rei01a} has shown that Knill's proposal is actually closely related to the aforemetioned quantum Reed-Muller code \cite{BraKit05}.  All of these protocols are what we call $n$-to-$1$ qubit stabilizer reductions:
\begin{definition}
\label{def:stabilizerRED}  An \textbf{n-to-1 qubit stabilizer reduction} performs the following:  (i) take an $n$-qubit resource $\rho_{\mathrm{nR}} $; (ii) measure the $(n-1)$ generating operators of an $n$-qubit stabilizer code $\mathcal{S}_{n-1}$ with one logical qubit; (iii) postselect on the all ``+1" measurement syndrome; (iv)  decode the logical qubit of the stabilizer code code onto a single output qubit, $\rho'$.
\end{definition}

When we obtain the desired measurement outcome, the resource state is projected into the codespace of the stabilizer code supporting one logical qubit.  All states inside the codespace are stabilized, $s\ket{\psi}=\ket{\psi}$, by all operators $s$ that are elements to the stabilizer of the code $\mathcal{S}_{n-1}$.  An encoded basis within this codespace is defined by a logical Pauli operator $X_{L}$  that is not an element of the stabilizer code but does commute with all elements of the stabilizer.  The encoded basis states are then $\ket{+_{L}}$ and $\ket{-_{L}}$, where $X_{L}\ket{\pm_{L}}= \pm \ket{\pm_{L}}$ and $s\ket{\pm_{L}}=\ket{\pm_{L}}$.  The projection operator onto the codespace can be expressed in terms of the encoded basis states:
\begin{eqnarray}
	P_{\mathcal{S}_{n-1}} = \kb{+_{L}}{+_{L}} + \kb{-_{L}}{-_{L}} \enspace .
\end{eqnarray} 
Another integral component of our definition of stabilizer reductions is \textit{decoding}.  We consider a decoding to be any Clifford unitary that maps $\ket{+_{L}} \rightarrow \ket{+}\ket{\phi} $ and $\ket{-_{L}} \rightarrow \ket{-}\ket{\phi} $, where $\ket{\phi}$ is a stabilizer state for all qubits except qubit 1.  Note that there is are many suitable decoding Clifford unitaries, partly because there are many suitable $\ket{\phi} $, but also because we are not interested in how the unitary maps states other than the encoded states $\ket{\pm_{L}}$.

Having defined the bulding blocks of a stabilizer reduction, it follows that after a successful stabilizer reduction the output qubit will be:
\begin{equation}
\label{eqn:StabOp}
 \rho' = \frac{\tr_{\mathrm{nR-1}} \left( C_{\mathrm{decode}}  P_{\mathcal{S}_{n-1}}  \rho_{\mathrm{nR}}   P_{\mathcal{S}_{n-1}}  C_{\mathrm{decode}}^{\dagger} \right) }{\tr \left(  C_{\mathrm{decode}}  P_{\mathcal{S}_{n-1}}  \rho_{\mathrm{nR}}   P_{\mathcal{S}_{n-1}}  C_{\mathrm{decode}}^{\dagger}  \right)} \enspace , 
\end{equation}
This is also a Clifford reduction with a single Kraus operator:
\begin{eqnarray} 
K & = &C_{\mathrm{decode}}  P_{\mathcal{S}_{n-1}}\enspace , \\ \nonumber
   & = & \kb{ +, \phi  }{+_{L}} + \kb{  -, \phi }{-_{L}}\enspace ,
\end{eqnarray}   
Whilst all stabilizer reductions are Clifford reductions, the converse is not true.

\section{Theorem Outline} \label{main}

Since the class of Clifford reductions encompasses the class of stabilizer reductions, one might be tempted to hypothesize that in some scenarios it is desirable to employ a Clifford reduction not based on stabilizer codes.  However, there is no existing evidence for this, as all existing protocols are essentially stabilizer reductions.  We say ``essentially" because a word of caution is necessary.  Many existing proposals make use of the idea of twirling \cite{BraKit05}, a randomizing process, which thus requires more than one Kraus operator for its description.  However,  while twirling is useful as an analytic tool for simplifying proofs, it is not necessary in a physical implementation.  For every stabilizer reduction that is preceded by twirling, there exists a derandomized stabilizer reduction where one applies the optimal choice of Clifford unitaries.   This will be amongst the results we prove in deriving our main theorem, which we now state: 

\begin{theorem} 
\label{THMreduction}
For all $n$-to-$1$ qubit Clifford reductions, all $n$-qubit resources $\rho_{\mathrm{nR}} $ and all single qubit pure states $\ket{\Psi}$, with an output qubit $\rho'$ of fidelity $\bra{\Psi}\rho'\ket{\Psi}$, at least one of the following is true:
\begin{enumerate}
\item[(i)] there exists a stabilizer state with an equal, or greater, fidelity, so $\bra{\Psi}\rho'\ket{\Psi} \leq |\bk{\Psi}{\phi}|^{2}$, where $\ket{\phi}$ is any single qubit pure stabilizer state; \textit{or} \\
\item[(ii)] there exists an $n$-to-$1$ stabilizer code reduction that also consumes $\rho_{\mathrm{nR}} $ and outputs a qubit $\rho''$ with equal, or greater, fidelity, such that  $\bra{\Psi}\rho'\ket{\Psi}  \leq \bra{\Psi}\rho''\ket{\Psi} $. 
\end{enumerate}
\end{theorem}

Our theorem contains two clauses, later to referred as clause \textit{(i)} and \textit{(ii)}, and either one or both clauses may hold true in any individual case.  However, only clause \textit{(ii)} is interesting in the context of magic state distillation, as when clause \textit{(i)} holds we could have achieved the same functionality without consuming the resource state $\rho_{\mathrm{nR}} $  at all.  


\section{Clifford Kraus Operators}
\label{sec:CliffordKraus}

Before proceeding, we derive a canonical form for Clifford group Kraus operators that we will employ throughout this paper. By definition, Clifford group operators conjugate Pauli operators to Pauli operators, such that $C \sigma' C^{\dagger}=\sigma$ and  hence $\sigma C=C \sigma' $.  Since Pauli projectors are composed of a Pauli operator and the identity ---  recall $P=(\unity+\sigma)/2$  --- then we also find that $ P C = C P' $.  All Clifford Kraus operators are composed of some sequence of Clifford unitaries and Pauli projectors, but by repeated conjugation all Clifford unitaries can be brought to the end of the operator, such that:
\begin{eqnarray}
	K & = & k C_{N}....C_{2}C_{1}P'_{N}...P'_{2}P'_{1} \enspace ,
\end{eqnarray}
where we drop the subscript of $K_{i}$ for brevity.   Next, we prove that sequential non-commuting Pauli projectors, $P'_{i}P'_{i+1}$,  can be simplified to a single Pauli projector.  Each of these Pauli projectors has an associated stabilizer $s'_{i}$ and $s'_{i+1}$, which must anticommute when the projectors do not commute.   Therefore:
\begin{eqnarray}
  	P'_{i}P'_{i+1} & = & (\unity + s'_{i})(\unity + s'_{i+1})/4 \enspace , \\ \nonumber
			 & = & (\unity+s'_{i} + s'_{i+1}+ s'_{i} s'_{i+1}))/4 \enspace ,\\ \nonumber
			  & = & (s'_{i} + s'_{i+1})(\unity+  s'_{i+1}))/4 \enspace , \\ \nonumber
			  & = & C' P'_{i+1} /\sqrt{2} \enspace ,
\end{eqnarray}
where $C'=(s'_{i} + s'_{i+1})/\sqrt{2}$ is a Clifford unitary that maps the projected subspace of $P'_{i+1} $ to that of $P'_{i}$ .  This shows that sequential anti-commuting projectors can always be replaced by a single projector followed by a Clifford unitary.  Continuing the process of conjugating through Clifford unitaries and removing non-commuting projectors, all $K$ can be reordered to begin with commuting stabilizer projections followed by Clifford unitaries:  
\begin{eqnarray}
\label{eqn:KrausSimple}
	K & = & k' C P  \enspace , 
\end{eqnarray}
with $P$ being:
\begin{eqnarray}
	P =  P''_{l}...P''_{2}P''_{1} & ; &  P''_{a}P''_{b}-P''_{b}P''_{a}=0 \enspace ;
\end{eqnarray}
a collection of stabilizer projectors associated with a codespace of some stabilizer code, $\mathcal{S}_{l}$, with $l$ generators ($s_{1}, s_{2},... s_{l}$) where $s_{j}=2P_{j}-\unity $.  

\section{Branch Expansion/Reduction}

Here we eliminate the need for considering many branches, and hence many $K_{i}$, by showing how one branch always provides an upperbound on the output fidelity.  However, before eliminating excess branches, we must unpack some hidden branches from~(\ref{eqn:WholeOp}).  This unpacking uses the property that a joint Hilbert space $\mathcal{H}\otimes \mathcal{H}'$, partially traced over Hilbert space $\mathcal{H}'$ satisfies:
\begin{equation} 
	\tr_{\mathcal{H}'} [  \rho ] = \sum_{j} \tr_{\mathcal{H}'} [ (\unity_{\mathcal{H}} \otimes \kb{\phi_{j}}{\phi_{j}} )\rho(\unity_{\mathcal{H}} \otimes \kb{\phi_{j}}{\phi_{j}})    ,
\end{equation}
where the set $\{ \ket{\phi_{j}} \}$ form any orthonormal basis over Hilbert space $\mathcal{H}'$.  This equation expresses the property that if one depolarises the state of an ancillary sub-system before tracing it out, it does not change the reduced state of the remaining sub-system.
Applying this formula to~(\ref{eqn:WholeOp}) gives:
\begin{equation}
 \rho'  \propto \tr_{\mathrm{nR-1},\mathrm{A}}  \left[  \sum_{i, \vec{j}}(\unity_{1} \otimes \kb{\vec{j}}{\vec{j}}) K_{i} \left( \rho_{\mathrm{nR}} \otimes  \kb{0}{0}^{\otimes m} \right)  K_{i}^{\dagger} (\unity_{1} \otimes \kb{\vec{j}} {\vec{j}} ) \right] \enspace ,
\end{equation}
where the bit-string $\vec{j}=( j_{2}, j_{3},.... j_{n+m} )$ has $n+m-1$  elements and specifies a computational basis state $\ket{\vec{j}}=\ket{ j_{2}, j_{3},.. j_{n+m}}$,  and since the partial trace excludes only the first qubit, this is the only qubit not included in $\ket{\vec{j}}$.  The summation is performed over all $i$ and all bit-strings $\vec{j}$.  Note that, for brevity we drop the denominator and hence use a proportionality sign.  We can now simplify the expression:
\begin{equation}
 \rho'  \propto \tr_{\mathrm{nR-1},\mathrm{A}}  \left[  \sum_{i, \vec{j}}  K_{i,  \vec{j}} \left( \rho_{\mathrm{nR}} \otimes \unity_{\mathrm{A}} \right)  K_{i, \vec{j}}^{\dagger} \right] \enspace ,
\end{equation}
where the new Kraus operators have incorporated the projectors:
\begin{equation}
\label{eqn:Kraus}
 K_{i,  \vec{j}}  = (\unity_{1} \otimes \kb{\vec{j}}{\vec{j}}_{\mathrm{nR-1},\mathrm{A}}) K_{i}  ( \unity_{\mathrm{nR}} \otimes \kb{0}{0}^{\otimes m}_{\mathrm{A}}) \enspace . 
\end{equation}
We say this unpacks hidden branches because we are now summing over two indices $i, \vec{j}$.  On first inspection, our introduction of these extra branches might seem mysterious, so we need some intuition to ground our mathematics.  We observe that the new Kraus operator $K_{i,  \vec{j}}$ always outputs a separable state,  as qubits  2 through to $n+m$ are all projected into a computational basis state, and so are disentangled from the output qubit, labeled 1.  In contrast, the original Kraus operator $K_{i}$ allowed for qubit one to be entangled with the other qubits, and so after after applying the partial trace we may incur further mixing of the output qubit.  In exposing these hidden branches, what we have gained is certainly of the state of all qubits except the output qubit.  

Having laid bare all branches, we now perform branch reduction, which relies on the convexity of fidelity measures.  In general, for any mixture $\rho' = \sum_{i} p_{i}\rho'_{i}$, the fidelity w.r.t $\ket{\Psi}$ is convex, and so:
\begin{equation}
	\bra{\Psi} \rho' \ket{\Psi} \leq  \bra{\Psi} \rho'_{\mathrm{Max}} \ket{\Psi} = \mathrm{Max}_{\rho'_{i}} \bra{\Psi} \rho'_{i} \ket{\Psi} \enspace .
\end{equation}
Hence, for any Clifford reduction and any $\ket{\Psi}$, the fidelity is upper-bounded by the fidelity produced by one of the branches, so that
\begin{equation}
\label{eqn:inequality1}
	 \bra{\Psi} \rho' \ket{\Psi}  \leq   \bra{\Psi}  \frac{ \tr_{\mathrm{nR-1},\mathrm{A}}  \left[   K_{\mathrm{Max}} \left( \rho_{\mathrm{nR}}\otimes \unity_{\mathrm{A}} \right)  K_{\mathrm{Max}}^{\dagger} \right]  }{ \tr \left[   K_{\mathrm{Max}} \left( \rho_{\mathrm{nR}}\otimes \unity_{\mathrm{A}} \right)  K_{\mathrm{Max}}^{\dagger} \right]} \ket{\Psi} \enspace ,
\end{equation}
where $ K_{\mathrm{Max}} = K_{\tilde{i},\tilde{\vec{j}}}$ is the Kraus operator with $i$ and $\vec{j}$ that maximize fidelity.  The validity of this expression can be conveyed by a simple metaphor about a foolish handyman and a smart handyman.  The foolish handyman has a toolbox containing a number of screwdrivers of varying quality, and whenever he wishes to use a screwdriver he reaches in and selects one at random, hoping it is the best.  The smart handyman has the same selection of screwdrivers, and also selects from his toolbox at random, however he only keeps his best screwdriver in his toolbox and leaves the rest at home.  Here we do the same with our best Kraus operator.  


\section{Exposing the Decoding}

Here we rearrange the optimal Kraus operator into a more intuitive form, which shows when the operator is either a trivial projection onto a stabilizer state or a more useful stabilizer reduction.  From~(\ref{eqn:Kraus}), we have:
\begin{equation}
 K_{\mathrm{Max}}  = (\unity_{1} \otimes \kb{\tilde{\vec{j}}}{\tilde{\vec{j}}}_{\mathrm{nR-1},\mathrm{A}}) K_{\tilde{i}}  ( \unity_{\mathrm{nR}} \otimes \kb{0}{0}^{\otimes m}_{\mathrm{A}}) \enspace . 
\end{equation}
Examining the last projector applied, $(\unity_{1} \otimes \kb{\tilde{\vec{j}}}{\tilde{\vec{j}}}_{\mathrm{nR-1},\mathrm{A}})$, this projects all but one qubit into a computational basis state.  Consequentially, the whole operator must project onto either a two-dimensional, logical, subspace or a definite stabilizer state.  In the former case, since all qubits but qubit 1 are left in a computational basis state, any logical qubit must have been decoded onto qubit 1.  This is the rough intuition for what we shall now show formally. We first expand out $K_{\tilde{i}}$ in the decomposition of~(\ref{eqn:KrausSimple}):
\begin{equation}
 K_{\mathrm{Max}}  = (\unity_{1} \otimes \kb{\tilde{\vec{j}}}{\tilde{\vec{j}}}_{\mathrm{nR-1},\mathrm{A}}) k_{\tilde{i}}C_{\tilde{i}}P_{\tilde{i}}  ( \unity_{\mathrm{nR}} \otimes \kb{0}{0}^{\otimes m}_{\mathrm{A}} ) \enspace . 
\end{equation}
This operator begins with two consecutive projections,  first $( \unity_{\mathrm{nR}} \otimes \kb{0}{0}^{\otimes m}_{\mathrm{A}}) $ then $P_{\tilde{i}}$.  These projectors always reduce to a single projector (see Sect.~\ref{sec:CliffordKraus}), possibly followed by some Clifford unitary:
\begin{equation}
 K_{\mathrm{Max}}  \propto  (\unity_{1} \otimes \kb{\tilde{\vec{j}}}{\tilde{\vec{j}}}_{\mathrm{nR-1},\mathrm{A}}) C_{\tilde{i}}'   ( P_{\mathrm{nR}} \otimes \kb{0}{0}^{\otimes m}_{\mathrm{A}}) \enspace . 
\end{equation}
where the first projector applied,  $\kb{0}{0}^{\otimes m}_{\mathrm{A}}$, must always remain the first applied, though it is supplemented by a possible extra projector, $P_{\mathrm{nR}}$, acting on the resource Hilbert space.  Next, we expand out the identity operation, such that:
\begin{equation}
	\unity_{1} = \kb{+}{+}_{1}+ \kb{-}{-}_{1} \enspace ,
\end{equation}
We make this substitution to track how our potential logical qubit is influenced by the rest of the operator.  It follows that:
\begin{eqnarray}
 K_{\mathrm{Max}}  & \propto &  (  \kb{+, \tilde{ \vec{j} } }{+, \tilde{ {\vec{j}}} }+ \kb{-, \tilde{ {\vec{j}}}}{-, \tilde{ {\vec{j}}}}  ) C_{\tilde{i}}'   ( P_{\mathrm{nR}} \otimes \kb{0}{0}^{\otimes m}_{\mathrm{A}}) \enspace .
\end{eqnarray} 
Next, we absorb the Clifford unitary into these stabilizer states, so that:
\begin{eqnarray}
\label{eqn:PreFormKraus}
 K_{\mathrm{Max}}   & \propto  & (  \kb{+, \tilde{\vec{j}}}{+_{L}} + \kb{-, \tilde{\vec{j}}}{-_{L}}  )   P_{\mathrm{nR}} \otimes \kb{0}{0}^{\otimes m}_{\mathrm{A}} \enspace , 
\end{eqnarray}
where $\ket{\pm_{L}}=(C_{\tilde{i}}')^{\dagger}\ket{\pm, \tilde{\vec{j}}}$ are orthogonal stabilizer states.  All that remains is to determine the effect of the projector $P_{\mathrm{nR}} \otimes \kb{0}{0}^{\otimes m}_{\mathrm{A}} $ on these stabilizer states, and one can verify (details in App.~\ref{App}) that this projection gives either:
\begin{enumerate}
\item[\textit{(a)}]  $ K_{\mathrm{Max}}    \propto   \kb{\phi_{1}, \tilde{ \vec{j} }  }{\pm'_{L}}  )  $ where $\ket{\phi_{1}}$ is a single qubit stabilizer state on qubit 1; or \\
\item[\textit{(b)}]  $ K_{\mathrm{Max}}    \propto   \kb{+, \tilde{ \vec{j} }  }{+'_{L}} + \kb{-, \tilde{ \vec{j} } }{-'_{L}}  )  $ where $\ket{\pm'_{L}}$ are orthogonal stabilizer states.
\end{enumerate}
\noindent
These different forms correspond to clause \textit{(i)} and \textit{(ii)} of our theorem. For form \textit{(a)}, the maximized Kraus operator $K_{\mathrm{Max}}$ projects qubit one onto a stabilizer state $\ket{\phi_{1}}$, and this state gives an upper bound on the fidelity of the Clifford reduction.  Hence clause \textit{(i)} of the theorem is always true when $K_{\mathrm{Max}}$ has form \textit{(a)}.  Although form \textit{(b)} is a stabilizer reduction, it still acts on the ancillary Hilbert space.  In the next section, we show that the ancillary component can be eliminated, and so form \textit{(b)} will entail the truth of clause \textit{(ii)} of our theorem.

\section{Ancillae Add Nothing}

Here we show that when $K_{\mathrm{Max}}$ has form \textit{(b)}, we can find an equivalent Kraus operator that acts on only the resource qubits but outputs the same mixture.  To see this, we first note that because  $(P_{\mathrm{nR}} \otimes \kb{0}{0}^{\otimes m}_{\mathrm{A}} ) \ket{\pm'_{L}}=\ket{\pm'_{L}}$, we can conclude that $\ket{\pm'_{L}}=\ket{\pm''_{L}}_{\mathrm{nR}} \ket{0}^{\otimes m}_{\mathrm{A}}$.  Furthermore, we also know that $\ket{\pm, {\tilde{\vec{j}}}}$ is a separable state such that $\ket{\pm, {\tilde{\vec{j}}}}=\ket{\pm,{\tilde{\vec{j}}_{\mathrm{nR}}}}_{nR}\ket{\tilde{\vec{j}}_{\mathrm{A}}}_{\mathrm{A}}$.  Therefore, the Kraus operator has a simple tensor product structure w.r.t the ancillary/resource partitioning:
\begin{equation}
\label{eqn:KmaxAncilla}
  K_{\mathrm{Max}}   \propto \mathcal{K}   \otimes \kb{\tilde{\vec{j}}_{\mathrm{A}}}{0}^{\otimes m}  \enspace  ;  
\end{equation}
 where we introduce the calligraphic $\mathcal{K}$ to describe the Kraus operator that only acts on the resource Hilbert space:
 \begin{equation}
   \mathcal{K}  =  \kb{+, \tilde{\vec{j}}_{\mathrm{nR}}}{+''_{L}} + \kb{-, \tilde{\vec{j}}_{\mathrm{nR}}}{-''_{L}} \enspace .
 \end{equation}
Substituting~(\ref{eqn:KmaxAncilla}) into~(\ref{eqn:inequality1}), we see that the ancillary terms can be traced out:
\begin{eqnarray}
\label{eqn:inequalitytraced}
	 \bra{\Psi} \rho' \ket{\Psi} &  \leq &   \bra{\Psi}  \frac{ \tr_{\mathrm{nR-1},\mathrm{A}}  \left[  (\mathcal{K}   \otimes \kb{\vec{j}_{\mathrm{A}}}{0}^{\otimes  m}) \left( \rho_{\mathrm{nR}}\otimes \unity_{\mathrm{A}} \right) (\mathcal{K}   \otimes \kb{\vec{j}_{\mathrm{A}}}{0}^{\otimes  m})^{\dagger} \right]  }{ \tr \left[   (\mathcal{K}   \otimes \kb{\vec{j}_{\mathrm{A}}}{0}^{\otimes m}) \left( \rho_{\mathrm{nR}}\otimes \unity_{\mathrm{A}} \right)  (\mathcal{K}   \otimes \kb{\vec{j}_{\mathrm{A}}}{0}^{\otimes  m})^{\dagger} \right]} \ket{\Psi} \enspace , \\ \nonumber
		 \bra{\Psi} \rho' \ket{\Psi} &  \leq &   \bra{\Psi}  \frac{ \tr_{\mathrm{nR-1}} \left[  \mathcal{K}  \rho_{\mathrm{nR}}  \mathcal{K}^{\dagger}   \right] }{  \tr \left[  \mathcal{K}  \rho_{\mathrm{nR}}  \mathcal{K}^{\dagger}   \right] } \ket{\Psi} \enspace .
\end{eqnarray}
This expressions proves clause \textit{(ii)} of our theorem, as $\mathcal{K}$ can always be accomplished by an $n$-to-$1$ qubit stabilizer reduction.  

\section{Conclusions \& Acknowledgments}

We have proven a theorem that shows that any protocol that uses Clifford group operations to reduce a non-stabilizer resource to a single qubit has a fidelity upperbounded by the fidelity of either a trivial protocol or a protocol derived from a stabilizer code, which we call a stabilizer reduction.  Our theorem reveals why all known protocols for magic state distillation can be described by stabilizer reductions.  Furthermore,  this theorem will form an essential component in a following paper~\cite{CampBrowne09b} where the authors prove that for a particular family of resource states, all possible stabilizer reductions, and hence all possible Clifford reductions, do not have distillation thresholds that are tight against the set of stabilizer states.   

The proof presented here has been cast in a general framework that compares two different classes of protocols, but it seems likely that there are numerous possible extensions.  It should be fairly easy to extend the proof to cover qu\textit{d}it generalizations of the Clifford group and stabilizer states.  More speculative is the question of whether a similar theorem holds for continuous variable quantum devices, where gaussian states and gaussian operations are frequently analogous~\cite{Bart02} to stabilizer states and Clifford group operations.  Finally, it might also be worthwhile exploring whether entanglement distillation obeys a similar theorem when the ideal Clifford group operations are also restricted to be local with respect to some partitioning.

The authors would like to thank Shashank Virmani, Matthew Hoban, Tobias Osborne, Ben Reichardt and Steve Flammia  for interesting discussions.    We acknowledge support from the Royal Commission for the Exhibition of 1851, the QIP IRC, QNET and the National Research Foundation and Ministry of Education, Singapore.

\appendix

\section{Appendix: Deriving the Two Forms (a) and (b)}
\label{App}

Here we show that the Kraus operator of~(\ref{eqn:PreFormKraus}) must of either form \textit{(a)} or \textit{(b)}.   By definition Pauli projectors take stabilizer state to stabilizer states with the addition of some constant, so:
\begin{eqnarray}
 K_{\mathrm{Max}}   & \propto  &  c_{+}  \kb{+, \tilde{\vec{j}}}{+'_{L}} + c_{-} \kb{-, \tilde{\vec{j}}}{-'_{L}} \enspace , 
\end{eqnarray}
where $\ket{\pm'_{L}}=(  P_{\mathrm{nR}} \otimes \kb{0}{0}^{\otimes m}_{\mathrm{A}})\ket{\pm_{L}}/\sqrt{c_{\pm}}$ are stabilizer states but are not necessarily orthogonal.    The variables $c_{\pm}$ can be assumed to be real by absorbing any phase into the definitions of $\ket{\pm'_{L}}$. Since relative phase may prove important, we remark that the phase absorbed will always be a multiple of $i$, and we will account for its effect later.  If one term vanishes, either because $c_{+}=0$ or $c_{-}=0$, then:
\begin{eqnarray}
(a1):  K_{\mathrm{Max}}   & \propto  &   \kb{\pm, \tilde{\vec{j}}}{\pm'_{L}}   ;    \enspace , 
\end{eqnarray}
which we refer to as form \textit{(a1)} as this Kraus operator projects onto a stabilizer state, but in general projections onto a stabilizer state may occur in a different basis.  When neither term vanishes, the coeffecients must be equal, so $c_{+}=c_{-}$.  To deduce this equality, we first consider the first Pauli measurement, $P_{i}=(\unity+s_{i})/2$, on the stabilizer states $\ket{\pm'_{L}}$.  Because the stabilizing operators of $\ket{+'_{L}}$ and $\ket{-'_{L}}$  differ only in phase, either both states are eigenstates of $s_{i}$ or neither state is an eigenstate of $s_{i}$.  For eigenstates of a projector, the projector either has no effect, or maps the state to zero.  Hence, when  $c_{+} \neq 0$ or $c_{-} \neq 0$, the projectors acting on eigenstates must have no effect.  As for projectors where  $\ket{+'_{L}}$ and $\ket{-'_{L}}$ are not eigenstates of $s_{i}$, the measurement outcomes are equally random, so $\tr[P_{i}\kb{\pm_{L}}{\pm_{L}}]=1/2$.  The same argument holds for subsequent Pauli measurements, and so we have that  $c_{+}=c_{-}=(1/2)^{m}$ where $m$ is the number of component Pauli projections with a non-trivial effect.  Now, if $\ket{+'_{L}}$ and \ket{-'_{L}} retain their orthogonality, then form \textit{(b)} follows immediately:
\begin{eqnarray}
(b):  K_{\mathrm{Max}}   & \propto  &   \kb{+, \tilde{\vec{j}}}{+'_{L}} + \kb{-, \tilde{\vec{j}}}{-'_{L}} \enspace .
\end{eqnarray}
If we wish to remove any phase absorbed into the definition of $\ket{\pm'_{L}}$, this is simple as the phase is always a factor of $i$, and this can be removed by adding a subsequent Clifford unitary $\unity \otimes (\kb{+}{+} + i^{N}\kb{-}{-})_{1}$.

The last remaining possibility is when neither term vanishes but $\ket{+'_{L}}$ and $\ket{-'_{L}}$ lose their orthogonality.  Since the initial states had commuting stabilizing operators, and the projection can not change this, if the vectors $\ket{\pm'_{L}}$  are not orthogonal they must correspond to the same state up to a relative phase $i^{N}$, and so $\ket{-'_{L}}=i^{N}\ket{+'_{L}}$.  Hence, in this case
\begin{eqnarray}
(a2):  K_{\mathrm{Max}}   & \propto  &   \kb{+, \tilde{\vec{j}}}{+'_{L}} + i^{N}\kb{-, \tilde{\vec{j}}}{+'_{L}} \enspace , \\ \nonumber
   & \propto  &   \kb{\phi_{1}, \tilde{\vec{j}}}{+'_{L}} \enspace .
\end{eqnarray}
where $\bra{\phi_{1}}=(\bra{+}+i^{N}\bra{-})/\sqrt{2}$ is a stabilizer state on the equator of the Bloch sphere.  Hence, form (\textit{a1}) and form (\textit{a2}) account for all possible single qubit projections, and hence jointly comprise form (\textit{a}).  This completes our proof that we always have either form (\textit{a}) or (\textit{b}).

\end{document}